\documentstyle[twocolumn,prc,aps,epsfig]{revtex}
\input epsf
\newcommand{\be}{\begin{eqnarray}}
\newcommand{\ee}{\end{eqnarray}}

 \newcommand{\gsim}{\mathrel{\hbox{\rlap{\lower.55ex \hbox {$\sim$}}
                   \kern-.3em \raise.4ex \hbox{$>$}}}}
\newcommand{\lsim}{\mathrel{\hbox{\rlap{\lower.55ex \hbox {$\sim$}}
                   \kern-.3em \raise.4ex \hbox{$<$}}}}

\def\tl{\tilde l}

\def\roughly#1{\mathrel{\raise.3ex\hbox{$#1$\kern-.75em%
\lower1ex\hbox{$\sim$}}}}
\def\lsim{\roughly<}
\def\gsim{\roughly>}

\begin{document}

\twocolumn[\hsize\textwidth\columnwidth\hsize\csname @twocolumnfalse\endcsname

\title {Understanding the Strong Coupling Limit\\
of the ${\cal N}=4$ Supersymmetric Yang-Mills at Finite Temperature
  }
\author {Edward Shuryak and Ismail Zahed}
\address {Department of Physics and Astronomy, State University of New York,
Stony Brook, NY 11794}
\date{\today}
\maketitle
\begin{abstract}
Recently, a number of intriguing results have been obtained 
for strongly coupled ${\cal N}=4$ Supersymmetric Yang-Mills 
theory in vacuum and matter, using the AdS/CFT correspondence.
In this work, we provide a physical picture supporting and 
explaining most of these results within the gauge theory.
The modified Coulomb's law at strong coupling forces static 
charges to communicate via the high frequency modes of the 
gauge/scalar fields. Therefore, the interaction between even
relativistically moving charges can be approximated by a potential.
At strong coupling, WKB arguments yield a series of deeply bound states,
whereby the large Coulomb attraction is balanced by centrifugation.
The result is a constant density of light bound states at {\it any}
value of the strong coupling, explaining why the thermodynamics
and kinetics are coupling constant independent. In essence, at strong 
coupling the matter is not made of the original quasiparticles but of 
much lighter (binary) composites. A transition from weak to strong
coupling is reminiscent to a transition from high to low 
temperature in QCD.
We establish novel results for screening in vacuum and matter through 
a dominant set of diagrams some of which are in qualitative agreement 
with known strong coupling results.
\end{abstract}
\vspace{0.1in}
]
\newpage

\section{Introduction}\label{intro}

${\cal N}=4$ super-Yang-Mills (SYM) is the most famous example of 
a Conformal Field Theory (CFT) in 4 dimensions. This theory has zero 
beta function and a non-running coupling constant, which can be continuously
changed from weak to strong. Unlike QED or QCD where for a critical
coupling $\alpha\approx 1$ there is vacuum rearrangement, the CFT is
believed to remain in the same Coulomb (plasma-like) phase for all
couplings, even strong ones $\lambda=g^2 N_c\gg 1$. 
Thus, it provides an interesting theoretical laboratory for understanding
properties of a strongly coupled Quark-Gluon Plasma (QGP) in QCD, which occurs
in and around the critical temperature $T_c$, as discussed in our recent 
paper~\cite{SZ_newqgp}.

A key breakthrough in making the results for the strong coupling regime
within reach was the
AdS/CFT correspondense suggested by Maldacena~\cite{maldacena}.
This conjecture has turned the intricacies of strong coupling
gauge theories to a classical problem in gravity albeit in 10
dimensions, leading to the static heavy-quark potential
~\cite{maldacena,rey,Rey_etal}, small angle scattering~\cite{zahed,janik} and 
large angle scattering~\cite{polchinski}. 
For instance, the static potential between a heavy quark
and antiquark follows from a minimal surface (classical string) 
between the quarks streched by gravity (metric of the AdS
space) as depicted in Fig. 1a. The result is a modified Coulomb's 
law~\cite{maldacena,rey}.

\be \label{eqn_new_Coulomb}
V(L)= -{4\pi^2  \over \Gamma(1/4)^4 }{\sqrt{\lambda} \over  L}
\label{coulomb}
\ee
for $\lambda\gg 1$. The numerical coefficient in the first bracket is 0.228.
The latter will be compared to the result from a diagrammatic
resummation below.

\begin{figure}[h!]
\centering
\includegraphics[width=8cm]{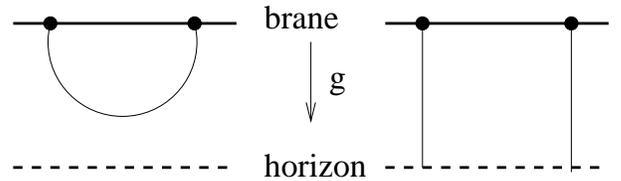}
\vskip .3cm
\caption{\label{fig_pbrane_Debye} 
Two types of solutions describing the potential between two static charges
(large dots) in the ordinary 4-d space (on the D3 brane). 
The string originating from them
can either connect them (a) or not (b). In both cases the string
is deflected by a background metric (the gravity force indicated by the
arrow marked g) downward, along the 5-th coordinate. After the string
touches the black hole horizon (b) a Debye screening of the interaction takes place.  
}
\end{figure}

The case of  non-zero temperature  is represented in the AdS
space by the occurence of a black-hole in the 5th dimension,
whereby its Schwarchild radius is identified with the inverse
temperature. When the string between charges
extends all the way to the black hole as shown in Fig. 1b, 
the heavy quark potential is totally screened for
a Debye radius of order $1/T$~\cite{rey}. This is to be contrasted with
$\sqrt{\lambda}T$ expected in the weak coupling limit for the electric
modes, and $\lambda\,T$ for the magnetic modes.

The AdS/CFT correspondence was also used to generate a number of
finite temperature results at strong coupling, including the free
energy~\cite{thermo}, the electric Debye screening~\cite{Rey_etal},
and the viscosity~\cite{PSS}. Also a number of real-time correlators 
were recently investigated, leading a tower of equi-distant but 
unstable resonances~\cite{Starinets,Teaney}. Their origin remains 
a mystery which we will attempt to explain.

{\it The main puzzle} related with all these results whether it 
is the free energy, the viscosity, or the resonance frequencies,
is their independence on the coupling $\lambda$ in strong coupling.
We recall that the interaction between the (quasi) particles
such as (\ref{eqn_new_Coulomb}) is proportional to $\sqrt{\lambda}$,
and the strong-couping Debye distance is $1/r\sim T$, so
the relevant quasiparticle energy scale must be $\sqrt{\lambda}T$. 
In a naive picture of  matter being
essentially a plasma of quasiparticles (weak coupling)
one would expect the interaction terms
of such order to show up in the free energy. 

{\it The main objective} of this paper is to explain this puzzle,
in the process of which novel results will also be derived. This
is achieved in two major steps as we now detail:

{\bf i.}  First, we attempt to understand the dynamical picture behind
the modified Coulomb law, and its Debye-screened form at finite $T$ and
strong coupling. For that, we identify  in the gauge theory a set
of diagrams whose resummation can reproduce the parametric
features of the above mentioned strong coupling results. As a result,
we learn an important lesson: in the strong coupling regime even the
static charges communicate with each other via high frequency
gluons and scalars,  propagating with an effective super-luminal velocity
$v\approx \lambda^{1/4}\gg 1$. 

{\bf ii.} Second, we argue that even for relativistically moving 
quasiparticles the interaction can be described by a
potential with a {\it near-instantaneous} and {\it  quasi-abelian} interaction,
a screened version of a modified Coulomb law given above.
Solving the Klein-Gordon (or Dirac or Yang-Mills) equations for scalars
(or spinors or gluons) in a WKB approximation yields {\it towers of deeply bound states},
extending from large quasiparticle masses $m/T\approx \sqrt{\lambda}$
all the way to small ones $E/T\approx\lambda^0$ that are independent
of the coupling constant. Their existence  in a strongly coupled
plasma at any value of the strong coupling, 
explains the main puzzle mentioned above. So, the ${\cal N}=4$ SYM at
finite temperature and strong coupling is {\it not} a gas of  
quasi-particles, but a gas of (much lighter) {\it composites}.
 We have found a ``precursor'' of such phenomenon in a 
near-critical Quark-Gluon plasma in QCD~\cite{SZ_newqgp}.

Before explaining these results in a technical way, let us first
explain our motivations and some intuitive ideas which were 
important to us in deriving them. Let us first mention  that the  
observation that  potential-type diagrams in Feynman gauge  reproduce
the strong coupling regime has been made  by Semenoff and Zarembo
\cite{zarembo}, who have shown that the ladder resummation works for the
circular Wilson loop and qualitatively explains the modified Coulomb law
at strong coupling and zero temperature.

Intuitively, the reason for a potential-like  regime stems from the
fact  that for $\lambda\gg 1$ the time between subsequent exchange of 
quanta is very short. The cost of the repulsive Coulomb energy becomes 
prohibitively large at strong coupling, forcing both charges to almost 
simultaneously change their colors, keeping them oriented in mutually 
the most attractive positions. These interactions are naturally ordered 
in time, justifying the use of ladder-type approximations.

We extend these observations in two important ways:

{\bf i.} First, we identify higher order diagrams which contribute equally
to the potential, and those which are subleading. Although we will not
attempt to sum them all, we argue that the distribution of  
gluons/scalars accompanying a pair of static charges look like a 
``quasi-string'', with a small transverse size of order
$L/\lambda^{1/4}$ in comparison to the length $L$.

{\bf ii.} Second, we show how Wilson loops in matter get
their strong coupling Coulomb's law from an abelianized Coulomb's law 
over time scales of the order of $L/\lambda^{1/4}$, using ladder
diagrams. The short time scale in matter is conditioned by the high
frequency modes which yield a quasi-static potential over
times of order $\sqrt{L/\omega_p}$ with $\omega_p\approx \sqrt{\lambda}T$ the 
typical plasmon frequency.


The bulk screening in the ground state can also be characterized by the dielectric constant
$\epsilon=1+4\pi\,\chi_e$. The electric susceptibility follows
from a simple Drude-Lorentz model where the dipole response 
follows from a high frequency harmonic oscillator

\be
\chi_e\approx \frac{{\bf n}\,\lambda}{m\,\omega^2}\approx
\frac{m^3\,g^2}{m\,(m^2\sqrt{\lambda})}\approx \sqrt{\lambda}
\label{c1}
\ee
where all the scales are set by the condensate value $m(\lambda)$ which
is seen to drop in (\ref{c1}). The dipole size $1/m$ is small.
Thus $\epsilon\approx\sqrt{\lambda}$ and the vacuum Coulomb's law is
$-{\lambda}/{\epsilon\,L} \approx -{\sqrt{\lambda}}/{L}$,
the result obtained in large $N_c$ and strong coupling.

\section{Deriving The Modified Coulomb's Law}

We start our discussion by reminding the
reader of the (Euclidean) derivation of the
standard Coulomb's law between two attractive and {abelian} 
static charges.  In the first quantized form  in Feynman gauge
one simply gets it from a 00-component of the photon (gluon) propagator

\be
V(L)= -{\lambda \over 4\pi^2}   \int_{-\infty}^{+\infty} \,{dt 
\over t^2+L^2}
\label{cou1}
\ee
where $t$ is the relative time separation between 
the two charges on their world lines. In the abelian case whether
at strong or weak coupling, the interaction takes place at { all}
time virtualities resulting into the standard  {instantaneous} 
Coulomb interaction with $V(L)\approx -\lambda/L$. The non-abelian 
modified Coulomb's law (\ref{coulomb}) is seen to follow from 
the abelian Coulomb's law (\ref{cou1}) whereby the relative time 
interval is much shorter and of the order $L/\sqrt{\lambda}\rightarrow 0$.

The effects of retardation are best captured in covariant gauges
as illustrated here.
At $T=0$ Feynman gauge treats the gluons and scalars on
equal footing, which makes the supersymmetric structure of the
underlying ${\cal N}=4$ SYM more transparent in diagrams.
 However, the observation itself should of course be gauge independent.
For instance, in Coulomb's gauge the same result
should follow from something like

\be
V(L) \approx
\int_{-\infty}^{+\infty}\,dt\,\frac{-\lambda}{L}\,\delta(\sqrt{\lambda}t)\,\,.
\label{cou11}
\ee
where the instantaneous time is rescaled by $\sqrt{\lambda}$ to account
for the delay. We have not worked out how this regime is explained in
other gauges.

\begin{figure}[h!]
\centering
\includegraphics[width=6cm]{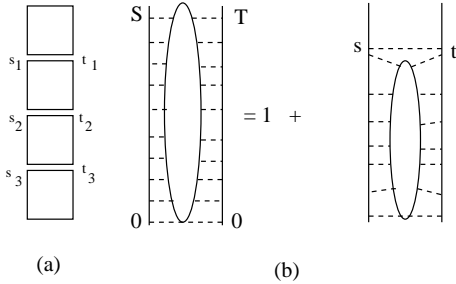}
\caption{\label{fig_ladders} 
(a) The color structure of ladder diagrams in the large-$N_c$
limit: each square is a different color trace, bringing the factor
$N_c$.
 The time goes vertically, and the planarity condition
 enforces strict time ordering, $s_1>s_2>s_3...$,
$t_1>t_2>t_3...$. (b) Schematic representation of the Bethe-Salpeter 
equation (\ref{eqn_BS}) summing ladders.
}
\end{figure}

\subsection{Summing Ladders}

We start by recalling some results established
in~\cite{zarembo} at zero temperature. At large $N_c$ 
(number of colors), the  diagrams can be viewed
as t'Hooft diagrams. An example of a ladder diagram
is shown in Fig.~\ref{fig_ladders}a, where the rungs can be either
gluons and scalars, as both are in the adjoint representation.
The first lesson is that each rung contributes a factor $N_c$,
which however only comes from planar diagrams. It means, that
in contrast to the Abelian theory, the time ordering should be strictly
enforced,  $s_1>s_2>s_3...$ and $t_1>t_2>t_3...$, which will be 
important below.

In Fig.~\ref{fig_ladders}b we schematically
depict  the Bethe-Salpeter equation. The oval connected with
the two Wilson lines by multiple gluon/scalar lines
is  the resummed Bethe-Salpeter kernel $\Gamma(s,t)$, describing
the evolution from time zero to times $s,t$ at two lines. It 
satisfies the following integral equation

\be \label{eqn_BS}
\Gamma({\cal S},{\cal T})=1+{\lambda \over 4\pi^2}\int_0^{\cal S} ds
\int_0^{\cal T} dt{1\over
    (s-t)^2+L^2} \Gamma(s,t) \ee
which provides  the resummation of all the
ladder diagram. $L$ is the distance between two charges, and 
the first factor under the integral is the (Euclidean) propagator for
one extra gluon/scalar added to the ladder. The kernel obviously
satisfies the boundary condition $\Gamma ({\cal S},0) =\Gamma(0,{\cal T})=1$.
If the equation is solved, the ladder-generated potential follows from

\be
V_{\rm lad}(L) 
=-\lim_{T\to{+\infty}}{\frac 1{\cal T} \Gamma\, ({\cal T},{\cal T})}\,\,,
\label{0a}
\ee

In weak coupling $\Gamma\approx 1$  and the integral on the rhs is
easily taken, resulting in 

\be
\Gamma ({\cal S,T}) \approx 1+\frac{\lambda}{8\pi}\,\frac{{\cal S+T}}L
\label{00a}
\ee
which results into the standard Coulomb's law.
Note that in this case the typical relative time difference
between emission and absorption of a quantum $|t-s|\approx L$, so one can
say that virtual quanta travel at a speed $v\approx 1$.

For solving it at any coupling, it is convenient
to switch to the differential equation

\be
\frac{\partial^2\Gamma}{\partial {\cal S}\,\partial {\cal T}} =
\,\frac{\lambda/4\pi^2}{({\cal S-T})^2+L^2}
\Gamma ({\cal S,T})\,\,\,.
\label{1a}
\ee
and change variables to
$x=({\cal S-T})/L$ and $y=({\cal S+T})/L$ through
\be
\Gamma (x,y) =\sum_{m}\,{\bf C}_m \gamma_m (x)\,e^{\omega_m y/2}
\label{2a}
\ee
with the corresponding boundary condition $\Gamma (x,|x|)=1$. The
dependence of the kernel $\Gamma$ on the relative times $x$ follows
from the differential equation

\be
\left(-\frac{d^2}{dx^2} -
\frac{\lambda/4\pi^2}{x^2+1} \right)
\,\gamma_m (x) = -\frac {\omega_m^2}{4}\,\gamma^m (x)
\label{3a}
\ee
For large $\lambda$ the dominant part of the potential in (\ref{3a})
is from {\it small} relative times $x$ resulting into a harmonic
equation~\cite{zarembo}

\be
&&\left(-\frac{d^2}{dx^2} +\frac 12
({\lambda/4\pi^2})\,x^2 \right)
\,\gamma_m (x) \nonumber\\
&&= -\frac 14 \left({\omega_m}^2-{\lambda}/{\pi^2}\right)\,
\gamma_m (x)\,\,.
\label{4a}
\ee
This shows that the sum of the ladders grow exponentially. At large 
times ${\cal T}$,  the kernel is dominated by the lowest harmonic mode of
(\ref{4a}). For large times ${\cal S\approx T}$ that is small $x$ and large 
$y$ 

\be
\Gamma (x,y)\approx {\bf C}_0\,e^{-\sqrt{\lambda}\,x^2/4\pi}\,
e^{\sqrt{\lambda}\,y/2\pi}\,\,.
\label{5a}
\ee
From (\ref{0a}) it follows that
in the strong coupling limit the ladder generated potential
is 

\be V_{\rm lad}(L)= -\frac{\sqrt{\lambda}/\pi}L \ee which 
has the same parametric form  as the one derived from the
AdS/CFT correspondence (\ref{eqn_new_Coulomb}) except for the
overall coefficient. Note that the difference
is not so large,  since $1/\pi=0.318$ is larger than the exact value  
0.228 by about 1/3. So additional screening,  left out to higher order
diagrams,  is needed 
to get it right.

\subsection{Higher order diagrams and a ``quasi-string'' regime }

The results of ref.  \cite{zarembo} discussed above indicate
that summing ladders get some vacuum physics but {\it not all} 
since the overall coefficient is not reproduced exactly. The same conclusion follows from
the fact that the expectation values of Wilson lines are gauge invariant, while
the ladder diagrams are not. Therefore, some non-ladder
diagrams must be equally important and should be included.

Before we discuss any higher order diagrams, let us summarize
what we have learned about the ladder resummation which are worth
stressing. For that it is convenient to return to the integral form 
of the BS equation (\ref{eqn_BS}) and note that the inclusion of the kernel
effectively forces the consecutive time steps between emission/absorption
moments  to be of the following duration~\footnote{
The time ordering of interaction points along each line,
 reflecting on the non-abelian character of the charges
at large number of colors is important here.}

\be \delta s\approx \delta t \approx L/\lambda^{1/4}\,\,.\ee
The overall number of time steps is huge and of the order of
$N\approx \lambda^{1/4}\,{\cal T}/L$ as can be seen by
the change of variables $(s,t)\rightarrow (x,y)$ in the
Bethe-Salpeter equation.
In the strong coupling regime the time steps are much
shorter than in weak coupling. This  implies
that the speed of propagation of {virtual} quanta
is forced to be parametrically {\it larger} than
the speed of light~\footnote{Since no information is carried over,
there is no problem with causality.}, with $v\approx \lambda^{1/4} \gg 1$. 
Thus it is the limitation on these two 
times which provides the reduction (or screening) factor
$1/\sqrt{\lambda}$  in Coulomb's Law for non-abelian gauge theories
at strong coupling.

\begin{figure}[h!]
\centering
\includegraphics[width=3cm]{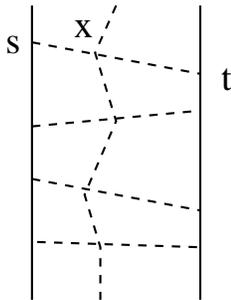}
\caption{\label{fig_extra_scalar} 
Examples of higher-order diagrams with an extra scalar/gluon connecting the
ladder rungs.
}
\end{figure}

Let us now look at higher order diagrams, e.g. 
Fig.~\ref{fig_extra_scalar} containing
extra scalar/gluon line connecting
the ladder rungs. Such diagrams can be resummed by a
modified Bethe-Salpeter equation, for the function
$\Gamma(s,t,{\bf x})$ where the last new argument is the
4d position of the new quantum on the rung.
The rhs of the new Bethe-Salpeter equation  has a modified
kernel, so the last term in (\ref{eqn_BS}) now  reads

\be 
\lambda^2 \int ds\, dt\,  d{\bf y}  
\Delta({\bf s}-{\bf y}) \Delta({\bf y}-{\bf t}) \Delta({\bf x}-{\bf y}) 
\,\Gamma(s,t,{\bf y})\,\,,
\label{high}
\ee
with ${\bf s}=(s,{\bf 0})$ and ${\bf t}=(t,{\bf L})$.
As before, the integral equation can be transformed into
a Schrodinger-like diffusion equation with a potential
defined by this kernel. Furthermore, in strong coupling
one can expand the denominators putting the small displacements
in the numerator. This result in now a Coulomb plus oscillator potential
of the type

\be
V\approx \lambda \left( \frac 1{({\bf y}-{\bf x})^2} +
\frac{{\bf C}}{L^2} + \frac{{\bf C}_{\mu\nu}}{L^4}\, {\bf y}_\mu {\bf y}_\nu\right)
\ee
where the 4d Coulomb~\footnote{In 4d the quadratic potential
does not create solutions falling to the center.} corresponds to the vertical 
propagator, from the old (${\bf y}$) to the new (${\bf x}$) position of
the extra zig-zag quantum in Fig.~\ref{fig_extra_scalar}.
Again, the resulting Schrodinger-like equation can be solved and
the lower level identified. Because of the large coefficient
$\lambda\gg 1$ the transverse steps would be as small as the time steps
namely,

\be \delta {\bf x}_t \sim
L/\lambda^{1/4}
\ee
Thus the $d^4{\bf x}$ integration yields a suppression factor of
$(1/\lambda^{1/4})^4\approx 1/\lambda$.  Since the $ds\,dt$ 
integrations bring about a suppression $(1/\lambda^{1/4})^2\approx
1/\sqrt{\lambda}$, it follows that (\ref{high}) is of order
$\sqrt{\lambda}$. This is precisely {\it the same order}
as the modified potential discussed in the previous subsection.
So, such higher order diagrams are neither
larger nor smaller than the ladder ones we discussed earlier.

\begin{figure}[h!]
\centering
\includegraphics[width=4cm]{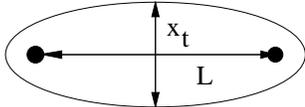}
\caption{\label{fig_planedistribution}
Distribution of the interaction vertices in space. The black
circles are static charges. 
}
\end{figure}

Needless to say, that one can also consider more elaborate ladder-type
sequences containing more than one extra quantum, with extra power
of the coupling again compensated by restrictions from extra $d{\bf y}$
integrations. Furthermore, since the number of quanta is not conserved
-- they can be absorbed by Wilson lines -- one should consider the extended
set of {\it coupled} BS equations for any number of quanta. We will not
attempt to write them down explicitly here. However, 
what we retain from this discussion, 
is that all extra quanta prefer to be in
an ellipsoidal region of space as shown
in Fig.~\ref{fig_planedistribution}.
The transverse sizes ${\bf x}_t\approx 
\delta t\approx L/\lambda^{1/4}$.

In QCD and other confining theories we known that gluons
make a string between two charges with a constant width and tension:
attempts to derive it from resummed diagrams continue. 
In CFT under consideration
the width in transverse direction must be proportional to $L$
since the conformal symmetry prohibits any other dimensional scale
to be developed. Still, in strong coupling
the ellipsoid is very elongated ${\bf x}_t\approx L/\lambda^{1/4}\ll L$:
this is what we meant by a ``quasi-string'' regime in the title of
this subsection.

In principle, the calculation based on the AdS/CFT correpondence
 not only yields the potential (\ref{eqn_new_Coulomb})
but also also the local distribution of the energy density in space 
as well. It would be interesting to investigate whether 
one can elevate the whole discussion from the global potential to the
differential distributions, in the form of resummed diagrams with extra 
quanta.

\section{Deeply bound states at finite $T$ }

\subsection{Bound states in the Quark-Gluon Plasma phase of QCD}

Before we discuss the occurence of bound states in the Coulomb
phase of CFT, we recall a similar occurence in non-supersymmetric
theories such as QCD in the deconfined or Coulomb phase. At high 
temperature the color charge in non-abelian gauge theories is 
deconfined but {\it screened} (rather than anti-screened as in  
the perturbative vacuum)~\cite{Shu_78}. The resulting phase
is called a Quark Gluon Plasma (QGP). Analytical and numerical 
(lattice) calculations concur on the fact that at very high
temperature, the fundamental fields (quarks and gluons for QCD)
in the QGP behave as free propagating quasiparticles~\footnote{Modulo color-magnetic
effects that are known to be non-perturbative at all temperatures~\cite{Pol_79}.}.

Although the perturbative series of the bulk thermodynamics, such as
the free energy, the pressure, the entropy, are found to be badly
divergent at $T$ comparable to $T_c$,
 it is somehow hoped that some sort of resummation
will make the weak-coupling quasi-particle picture work, as the
electric screening would keep the effective coupling weak for $T>T_c$.
The very first suggestions for the QGP signal were a disappearance of 
familiar hadronic peaks such as $\rho,\omega,\phi$ mesons
in the dilepton spectra \cite{Shu_QGP}. Even small-size deeply-bound
$\bar c c$ states such as $\eta_c,J/\psi$, were expected to melt in the QGP at
$T>T_c$~\cite{MS,KMS}. 

This picture has been challenged in our preceding work~\cite{SZ_newqgp},
where we argued that in the window of temperatures $T\approx$(1-3)$T_c$ 
the effective coupling can run up to large values. For  $\alpha_s\approx
1$ we have $\lambda=g^2 N_c\approx 40$, before it is cutoff by screening.
We had shown that this leads to (loosely) bound states of $\bar c
c,\bar q q, gg, qg, \bar{q}g$ s-wave states, explaining recent lattice 
observations~\cite{Karsch_charmonium,Karsch:2002wv}.   

This window of temperatures is very important, as it is
the only one which is experimentally accessible, e.g. by RHIC
experiments in Brookhaven National Laboratory. 
We argued  in our recent work~\cite{SZ_newqgp}
that the existence of such weakly bound states
leads to large scattering lengths and cross sections
of quasiparticle rescattering,
radically changing the kinetic properties of the QGP.
This observation is crucial for understanding why
in experiments, such as those at the Relativistic
Heavy Ion Collider, the QGP  behaves like a near-perfect liquid and
displays prompt collective hydrodynamical behavior.

\subsection{Heavy-light composites and related correlators}

In view of this recent development in QCD, we are led to ask whether
strongly coupled CFT in its Coulomb phase for all couplings (no phase
transition) allow for hadronic bound states, and what role if any 
those may play at finite temperature. To keep the presentation simple,
we will  exemplify our  method by analyzing the simplest ``heavy-light meson''
formed of a pair of an (auxiliary) static and light  fermions. As noted
by one of us~\cite{Shu_hl}, such states are the ``hydrogen atoms'' of
hadronic physics, allowing us to start with  a single-body problem
instead of a many-body one.

We focus on a static (infinitely heavy)
particle accompanied  by a light one with opposite (adjoint) color. 
In CFT the light particle can be either a scalar, a
gluino or a gluon~\footnote{The spin of the infinitely heavy charge is 
of course of no relevance and can be arbitrary.}. In general, one can
address the problem by  considering the correlator of a quark 
(antiquark) field with a time-like (heavy) Wilson line~\cite{Shu_hl} 
in Euclidean space,

\be
{\bf C} (T) =\left<{\rm Tr}\left(\Psi^+(T,{\bf 0})\,{\bf W} (T,0)\,
\Psi (0, {\bf 0})\,{\bf W}^+(T, {\bf 0})\right)\right>\,\,.
\label{cor}
\ee
with $\Psi$ the massless adjoint anti-quark of the SYM theory and 
\be
{\bf W} (T,{\bf 0}) ={\rm exp}\left(-ig\int_0^T\,dt\,\dot{x}(t)\cdot
A(t,{\bf 0})\right)
\ee
the infinitely heavy adjoint quark.
In the string-theory context this correlator would correspond
to an open string. It is possible to evaluate (\ref{cor}) 
using the AdS/CFT correspondence at finite temperature, i.e.
in the presence of a black hole. On the other hand, the correlator 
can be directly calculated by solving the Green function equation 
with a modified (strong coupling and screemed) Coulomb potential. If 
the two methods yield the same result, this would vindicate our
physical description based on a near-instantaneous potential.

Here instead, we note that the large Euclidean time behavior of the
correlation function (\ref{cor}) is controlled by the lowest heavy-light
fermion bound state, i.e ${\bf C}(T)\approx e^{-E\,T}$. The bound states
are solution to the relativistic Dirac (Klein-Gordon, Yang-Mills)
equation for gluinos (scalars, gluons) with a screened Coulomb 
potential to be derived below (see (\ref{Epot}) below). 
It is important to note that at finite temperature all quasi-particles
develop a mass (as well as a width through collisional broadening)
which sets the scale for the bound state problem even for the unscreened
Coulomb potential. In CFT the thermal masses are generically of the form

\be \label{eqn_m*}
m(T,\lambda)\approx T \lambda^\alpha \ee
with some power $\alpha$. Below we assume $\alpha=1/2$ in strong
coupling, based on the modified Coulomb law and Debye
length $1/T$. (The mass scale then happen to be
 exactly like in weak coupling.)

\subsection{Relativistic (WKB) Spectrum for the Coulomb potential}

The simplest bound state problem in our case is that of a gluino
in the presence of an infinitly heavy source with compensating color
charge. The latter acts as an overall attractive Coulomb potential
$V$ (the effects of screening will be discussed below). In strong
coupling $V$ acts on the accompanying relativistic gluino
quasi-instantaneously. For a spherically symmetric potential,
the Shrodinger-like equation is radial and reads

\be 
-{d^2   \over dr^2}\chi_l=\left[(E-V)^2-m^2-{\tl^2\over
r^2}\right]\chi_l
\label{ex}
\ee
with the gluino wavefunction $\phi=Y_{lm}\chi_l(r)/r$ and the orbital
quantum number $\tl^2=l(l+1)$. For a semiclassical analysis, we will
use the Langer prescription $l(l+1)\rightarrow (l+1/2)^2$, which is
known to yield semiclassically stable S-states. For a
Coulomb-like potential, (\ref{ex}) is exactly solvable in terms of
hyper-geometric functions, much like the nonrelativistic problem
for a hydrogen atom. It is however physically more transparent
to use a semiclassical treatment a-la-Bohr.

Identifying the lhs of (\ref{ex}) with the radial momentum
squared, one can readily construct the necessary ingredients of WKB. 
Those are the turning points, roots of the rhs of (\ref{ex}), which
for the generic Coulomb potential $V=-C/r$ are

\be \label{eqn_radii}
r_{1,2}= {1\over E^2-m^2}\left(EC\pm \sqrt{m^2 C^2+\tl^2(E^2-m^2)
}\right)
\label{ex1}
\ee
It is important that $r$ is a radial variable, so that both solutions
in (\ref{ex1}) are {\it positive}, for otherwise we are dealing with
either a scattering state returning to large $r$, or an inward falling
state with a wave towards small $r$. For small coupling $C$ the
positivity condition is always fulfilled, while for large $C$ 
the positivity condition is only fulfilled for orbital motion
with $\hat{l}\approx (\sqrt{\lambda}+1/\sqrt{\lambda})$.

Using the Bohr-Sommerfeld quantization one introduces the radial
quantum number and gets\footnote{In order to avoid confusion:
what we call $n$ is not the principal quantum number but just a radial
one. So there are no relation or limits on it for any $l$.}

\be 
&&\int_{r1}^{r2} p_r\, dr= \pi (n+1/2)\nonumber\\&&
=\pi\left(1-{ E_{nl}\over \sqrt{E_{nl}^2-m^2} }{C\over 
\sqrt{\tl^2-C^2}}\right)
\ee
from which the quantized levels are
 
\be \label{eqn_spectrum_strongcoupling}
E_{nl}=\pm m \left[1+
\left({C\over n+1/2+\sqrt{\tl^2-C^2} } \right)^2\right]^{-1/2}\,\,.
\ee
In weak coupling $C=g^2N=\lambda$ is small and the bound states energies
are close to $\pm m$. Specifically~\footnote{The fact that only the
combination $n+l$ appears, i.e. principle  quantum number, is a
consequence of the known Coulomb degeneracy. This is no longer
the case in the relativistic case.}

\be 
E_{nl}\pm m\approx -{C^2 m \over 2(n+l+1)^2}\,\,, 
\ee
which is the known Balmer formulae.
In the (opposite) strong coupling limit the coefficient is large
$C=(4\pi^2 / \Gamma(1/4)^4 )\sqrt{\lambda} \gg 1$. Unless the square
root gets balanced by a sufficiently large angular momentum,
the quantized energies are imaginary. This does not happen
for $n,\sqrt{\tl^2-C^2}\approx \lambda^0$ and $C\approx \sqrt{\lambda}$,
which are also the conditions for which both roots in (\ref{ex1}) are
positive. In this regime, one may ignore the $1$ in 
(\ref{eqn_spectrum_strongcoupling}) and obtain the {\it equi-distant}
spectrum

\be
E_{nl}\approx {m\over C}\left[ 
(n+1/2)+\left((l+1/2)^2-C^2\right)^{1/2} \right] 
\label{wkb}
\ee
Since the mass is related to the thermal loop with $m\approx
\sqrt{\lambda}$, the ratio $m/C\approx T\lambda^0$ is proportional to 
temperature but {\it  independent} of the coupling constant.

\begin{figure}[t!]
\centering
\includegraphics[width=5cm]{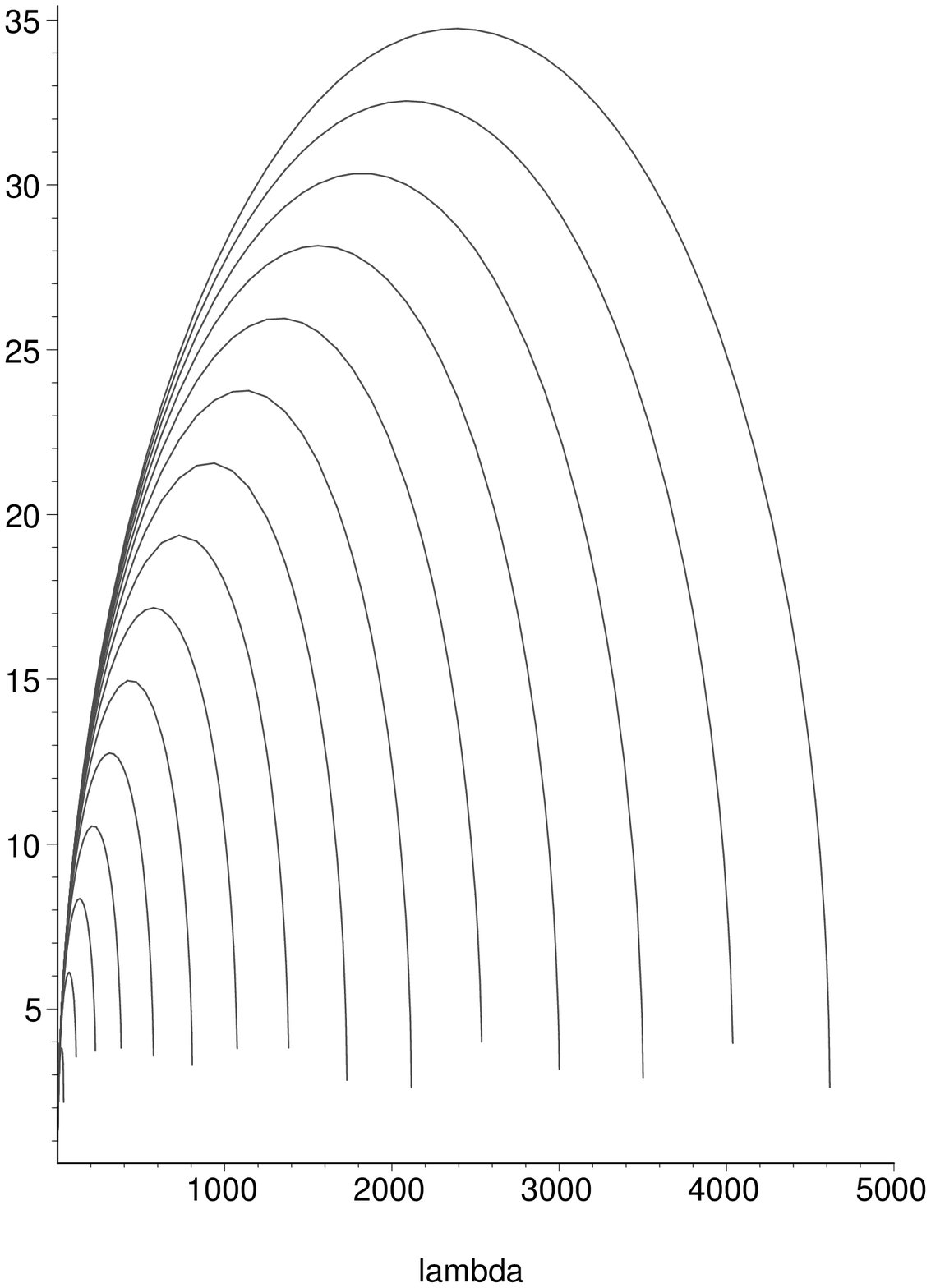}
\vskip .4cm
\includegraphics[width=5cm]{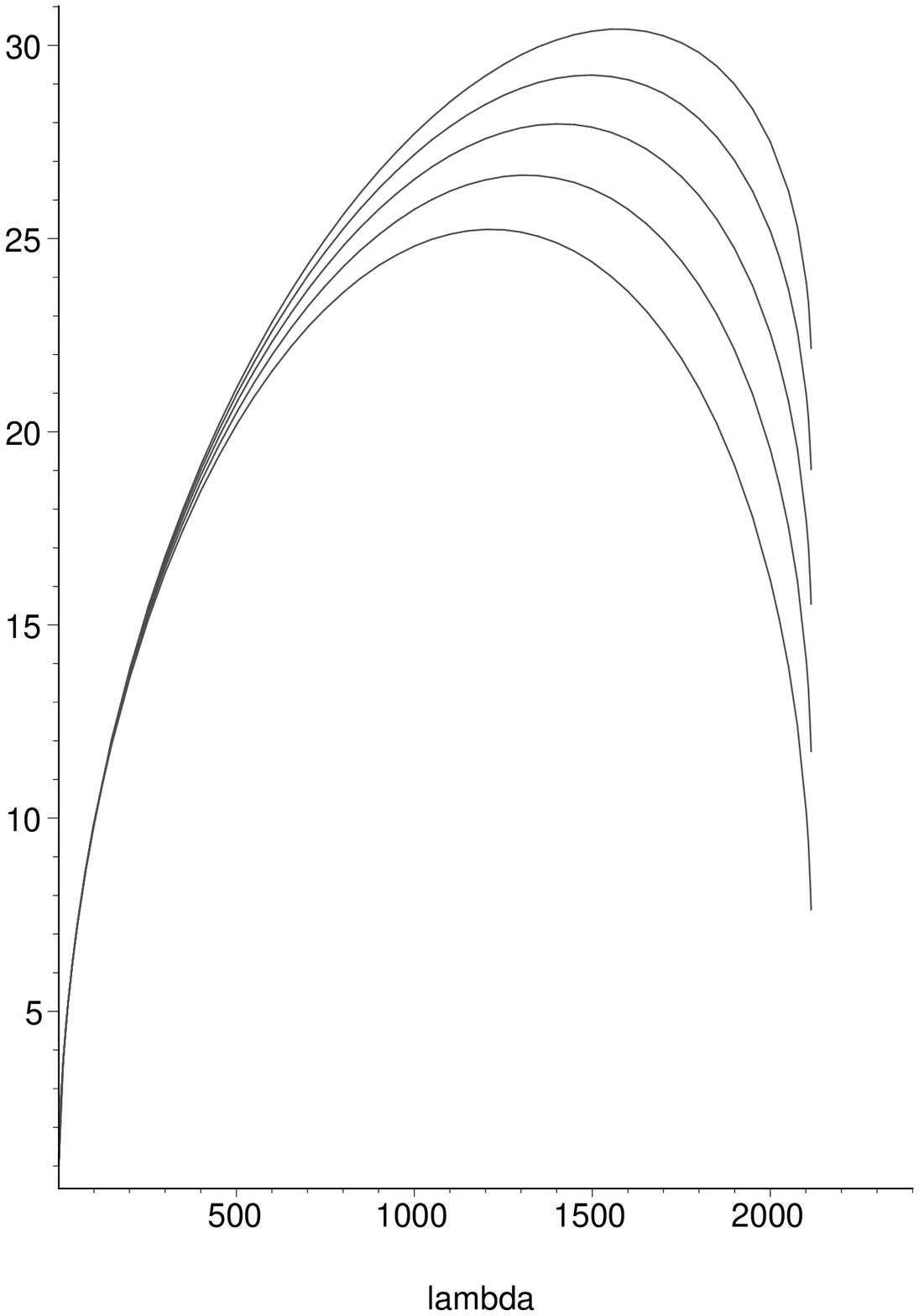}
\caption{\label{fig_wkb} 
The WKB spectrum versus 'tHooft coupling constant $\lambda$.
(a) levels with fixed n=0 and $l=1..15$, (b) levels with fixed
$l=10$ and $n=0..5$. 
}
\end{figure}

More details on the WKB spectrum are shown in Fig.~\ref{fig_wkb}
for different values of the orbital quantum number $l$ (a) and 
radial quantum number $n$ (b). All lines end when the Coulomb
attraction is able to overcome the centrifugal repulsion.

In summary we have put forward an explanation for the occurence
of a tower of light composite states with energies (masses) that
are independent of the strong coupling $\lambda$. The composites
are strongly bound Coulomb states which occur relativistically
due to a balance between the strong Coulomb attraction and the
repulsive centrifugation in an l-state. Only the states with
$(l+1/2)\ge\sqrt{\lambda}/\pi$ bind, resulting into an equidistant
WKB spectrum.  The critical coupling for S-states is

\be
g_c=\frac {\pi}{\sqrt{N_c}}\hspace{.35cm}{\rm or}\hspace{.35cm} \lambda_c=\pi^2 
\ee
Note that the
orbital states with $l\approx \sqrt{\lambda}$ are either bound (real)
or unbound (complex). The complex states correspond to the case were 
the light relativistic particle collapses onto the heavy source due to 
the large Coulomb attraction overcoming centrifugation. Above we have
speculated that at $T=0$ these states may pair condense in the form of
neutral dipoles at the origin of the large dielectric constant
$\epsilon\approx \sqrt{\lambda}$.

\subsection{Effects of screening}

In so far we have ignored in the WKB spectrum the effects of screening
on the Coulomb potential. To assess that, we will show below that in
strong coupling the screened Coulomb potential between scalars is mostly
mediated by the electric gluons, leading to 

\be
V({\bf x}) =- \frac{C}{|{\bf x}|}\,
\,{\bf F}_E(T{\bf x})\,\,
\label{propxx}
\ee
with 

\be
{\bf F}_E(T{\bf x}) =\frac{(\pi T|{\bf x}|)}{|{\rm sh}(\pi T|{\bf
x}|)|}
\label{Epot}
\ee
In strong coupling the effective screening
length is $1/(\pi T)$ independent of the coupling constant. 
This result is remarkably in agreement with lattice QCD 
simulations in the (1-3)$T_c$ window. The derivation
of (\ref{Epot}) using screened ladders graph with
$C=\sqrt{\lambda}/2\pi$ is given below. Note that 
at $T=0$ (\ref{propxx}) reduces to the unscreened Coulomb potential.

How would the screened potential (\ref{propxx}) affects our WKB
spectrum? The answer is only marginally. Indeed, if we insert 
(\ref{eqn_spectrum_strongcoupling}) back to the expressions for the radial turning
points (\ref{eqn_radii}), we find that the largest turning point 
occurs for $r_2\approx 1/T$ which is comparable to the screening
radius following from (\ref{propxx}). We note in passing, that the size
of the Coulomb bound state in strong coupling is very small and of order
$1/({\sqrt\lambda}T)$.

\subsection{More details on the bound states}

The semiclassical arguments developed above were curtailed to
scalars. How would they work for gluons and gluinos? To answer
qualitatively this question, we recall that in a weakly coupled 
plasma, the number and nature of the gluons and gluinos is different
from free massless waves of the free theory. For instance the
gluons carry 2 polarizations in free space, while in the plasma a 
third one appears, the plasmon~\cite{Shu_78}. Also the fermions
develope different excitation modes depending whether chirality
and helicity are the same or opposite~\footnote{We ignore the U(1) axial 
anomaly and instantons, and consider the $L,R$ chiralities to be
absolutely  conserved quantum numbers. }, see e.g.
\cite{weldon} for a general discussion and weak coupling
results. In the latter case the mode is called a plasmino.

If the dispersion laws for gluons and gluinos were known,
the effective equation of motion suitable for the discussion
of the bound state problem can be obtained by
standard substitution of the covariant
derivatives in the place of momentum and energy.
The electric effective potential would then go together with the
energy. For two gluinos, it is 

\be
V_{qQ}({\bf x}) =- \frac {C}{|{\bf x}|}\,
{\gamma_4}\,\,{\bf F}_E(T{\bf x})\,\,.
\label{propxxx}
\ee
Unfortunately, we do {\it not} know the pertinent dispersion laws for strongly
coupled CFT. We only know from lattice studies \cite{latt_quasipart_masses} that
in the window $T=$(1-3)$T_c$ for which the QCD plasma is likely in 
a moderatly strong Coulomb phase, the dispersion law for both quarks and gluons is 
of the canonical type $\omega^2=k^2+m_F^2$. For fermions, $m_F$ is not
the usual (chiral violating) mass term, but rather a thermal chiral mass.
This dispersion relation can be incorporated in a linear Dirac equation
through a chiral mass of the form $\gamma_4\,{m_F^2}/E$, resulting in 
the following equation

\be
\left(\left(E-\frac{m_F^2}{E}\right) +i\alpha\cdot\nabla +
\frac{\sqrt{\lambda}/2\pi}{x}\,{\bf F}_E(xT)\right)\,q({\bf x}) =0\,\,.
\label{SHR}
\ee
The above WKB analysis can be carried for (\ref{SHR}) as well, with
similar conclusions as the ones reached for the relativistic scalars.

\section{Thermodynamics and kinetics at strong coupling}

\subsection{AdS/CFT results}

The  CFT thermodynamics   
at strong coupling has been studied by a number of authors
\cite{thermo}. It was found that
the free energy in this limit is

\be \label{eqn_F}
{\bf F}(T,N_c,\lambda)=\left((3/4)+ O(1/\lambda^{3/2})\right)
{\bf F} (T,N_c, 0) \ee
where ${\bf F}(T,N_c,0)\approx N_c^2 T$ is the free (zero coupling) result,
analogous to the Stephan-Boltzmann result for blackbody radiation.
The kinetics of the finite-$T$ CFT at strong coupling was recently
investigated also using the AdS/CFT correspondence~\cite{PSS}. In
particular, the viscosity of strongly coupled matter was found to be
unusually small, leading to a rather good liquid with hydrodynamical
behavior even at small spatial scales.  In particular,  the sound 
attenuation length (analogous to the mean free path) was found to
be~\cite{PSS}

\be 
L_{sound}^{S}={4\over 3}{\eta \over s T}={1\over
 3\pi T}
\ee
where $\eta,s$ are the viscosity coefficient and stropy density, respectively. This length
 is much {\it shorter} than in the weak coupling   

\be 
L_{sound}^{W}=\frac 1{Tg^4 {\rm ln\,(1/g)}}\gg \frac 1T\,\,. 
\ee

As we have mentioned in the introduction, such results are rather 
puzzling dynamically from the gauge-theory standpoint. Normally all kinetic
quantities are related with scattering cross sections, and the absence
of {\it any} coupling is a priori implausible physically. In general,
thermodynamical quantities count degrees of freedoms and one may think 
that the coupling constant may indeed be absent. However, the effective 
masses of the quasiparticles and all pairwise interactions are still 
proportional to $\sqrt{\lambda}$  and in general $\lambda^\alpha$ with
$\alpha>0$. If so, the quasiparticle contributions to
the statistical sum should be exponentially small at strong coupling,
of the order of $ e^{-\lambda^\alpha}\ll 1$, and one will have to conclude that
some other degrees of freedom are at play.

Indeed there are new degrees of freedom at play in the strong coupling
regime of the thermal Coulomb phase. As we have
shown above, for large $\lambda$ there are light (deeply bound) binary
composites, with masses of order $T$ irrespective of how strong is
$\lambda$. The composites are almost point like with 
 thermal sizes of the order of $1/T$, which 
readily explains the liquid-like kinetic behaviour. The thermal Coulomb
phase is a liquid of such composites. The composites are light and should dominate the 
long-distance behavior of all the finite temperature Euclidean correlators.

\subsection{Gauge theory: Quantitatively}

We start by explaining the leading contribution (\ref{eqn_F}) from
the gauge theory point as a liquid of composites. The factor of $N_c^2$
in front of the free energy and the viscosity for instance, follows from
the fact that {\it all} our composites are in the adjoint
representation. Indeed, for fundamental charges the composites are
meson-like (color neutral) with all color factors absorbed in the
coupling constant. For adjoint charges, there is in addition {\it two}
spectator charges~\footnote{For example, a composite can be initially made of
blue-red and red-green colors, with red rapidly changing inside the
ladder. One can view it as two extra
vertical lines added to Fig.\ref{fig_ladders}(a).} 
that do not participate into the binding. The Coulomb phase
is not confining. Thus the number of light and Coulomb bound composites
is $N_c^2$.

The astute reader may raise the question that by now our arguments
are totally circular: we have started with $N_c^2$ massless relativistic
states and we have returned to $N_c^2$ light composites. So what is the
big deal? Well, the big deal is that we are in strong coupling, and in
fact the degrees of freedom completly reorganized themselves in
composites, for otherwise they become infinitly heavy thermally
and decouple. This is how Coulomb's law negotiates its deeds in a system
with very large charges.

The composites carry large angular momentum in strong coupling, 
i.e. $l\approx \sqrt{\lambda}$. Their weight contribution to the
partition function is of the order of

\be 
\int_{l_{min}}^{l_{max}} dl^2=l_{max}^2-l_{min}^2\approx \lambda^0
\label{lw}
\ee
which is independent of $\lambda$, since the WKB orbits are stable
only for $l\approx (\sqrt{\lambda} +1/\sqrt{\lambda})$. The ensuing
thermodynamical sum over the radial quantum number $n$ is
independent of $\lambda$, leading to a free energy ${\bf F} =-({\bf C}\pi^2/6)
N_c^2\,T$ in agreement with (\ref{eqn_F}). For weak coupling ${\bf C}=1$
while for strong coupling ${\bf C}=3/4$.

The overall coefficient in the $N_c^2$ part of the free energy in strong 
coupling is harder to obtain from the gauge theory side although the 
WKB spectrum could be used to assess it. A simple example on how this
arises in strong coupling and at large number of colors was shown in
\cite{hansson}. Alternatively, if we refer to
the coupling-valued coefficient by ${\bf C}(\lambda)$ and assume it to be an analytic 
function in the complex $\lambda$-plane modulo singularities, then 
($\lambda>0$) 

\be
{\bf C} (\lambda) ={\bf C} (0) +\frac
{\lambda}{\pi}\int^{\infty}_{+0}
d\lambda'\,\frac{{\rm Im}\,{\bf C} (-\lambda')}{\lambda'(\lambda'+\lambda)}\,\,.
\label{dis}
\ee
If the imaginary part along the negative real axis is bounded, 
the rhs is $\lambda$ independent in strong coupling. 
To proceed further requires a better
understanding of the singularity structure of the free energy in the
complex plane, e.g. the Dyson singularity in weak coupling which
suggests ${\rm Im}\,{\bf C} (-\lambda) \approx e^{-1/\lambda^\alpha}$ with
typically $\alpha=1$. This point will be pursued elsewhere.

\section{Screening at Finite Temperature: preliminaries}

At finite temperature and in Euclidean space, the Wilson loops can
be divided into electric (E) and magnetic (M) ones. the electric 
Wilson loop is sensitive to Coulomb's law between static charges,
while the magnetic Wilson loop is sensitive to Lenz's law between
two running currents. In the Euclidean vacuum $O(4)$ symmetry rotates
one to the other, and we expect the same modified Coulomb's law 
(\ref{0}) in CFT. In non CFT theories such as QCD they are both 
confining.

The usual definition of the ``screened potential'' implies that at 
$r\rightarrow \infty$ the potential vanishes. However, there is a
residual {\it negative} constant mass renormalization at finite $T$,
due to the self-interaction of a static charge with its thermal
cloud. In weak coupling it is of the order of $\lambda^{3/2}T$~\cite{Shu_78} 
using Debye's argument, and in strong coupling it is 
of the order of $\sqrt{\lambda}T$~\cite{Rey_etal}. This {\it negative}
correction to the mass of a static charge, should not be confused
with the {\it positive} effective mass of quasiparticles which
is of order $\sqrt{\lambda}T$ as well.

\subsection{Magnetic Screening and AdS/CFT}

At non-zero temperature $T$ Euclidean space is a cylinder of radius
$\beta=1/T$, and the $O(4)$ symmetry is reduced to $O(3)$ symmetry
with no a priori relationship between the electric and magnetic 
Wilson loops. In non CFT theories such as QCD and in weak coupling,
the electric charges are Debye screened while the magnetic charges
are not~\cite{Shu_78}. Weak-coupling non-perturbative effects related 
to spatial confinement in high temperature 3d YM theory, are believed to 
generate a magnetic mass of order $\lambda T$~\cite{Pol_79}. Several 
lattice calculations support this idea, leading a non-zero spatial string
tension of order $\lambda^2T^2$.

What happens at strong coupling? Our first new observation in CFT 
theories, is that at strong coupling the AdS/CFT correspondence
yields magnetic screening which is the {\it same} as the electric
screening observed in~\cite{Rey_etal}. Thus for large spatial
separation 

\be 
V_M (\beta ,L)  = V_E(\beta ,L) = {{\sqrt{\lambda}}\over L} f(\beta/ L)\,\,,
\ee
where $f(\beta/L)$ was evaluated in~\cite{Rey_etal} for electric Wilson
loops. The simple way to see this is to recall that the modified
Coulomb's law for the electric Wilson loop follows from the 
{\it pending} string in the 5th direction as shown in Fig. 2. 
This phenomenon is unmodified for spatial Wilson loops, whereby
the minimal surface is still of the same nature. Note that the
temperature in this case is given by a suitable choice of the
radius of the black hole. At strong coupling the electric and magnetic
scales are the same, since there is no apparent spatial confinement 
in high temperature CFT.

\subsection{Naive (Unscreened) Ladders at finite T}

Naively, the first step toward the calculation of the potential
between charges at finite temperature is
the same resummation of
the ladder diagrams via Bethe-Salpeter equation,
but with thermal propagators for gluons/scalars.
Let us see, for pedagogical reasons, what this
procedure will produce.  We can
apparently have either electric or magnetic ladders, thus the expression
becomes

\be
V_{E,M}(L) 
=-\lim_{{\cal T}\to{+\infty}}{\frac 1{\cal T}\, {\rm ln}\Gamma_{E,M}
({\cal T,T})}\,\,.
\label{0}
\ee
The electric kernel is generated through

\be
\frac{\partial^2\Gamma_E}{\partial {\cal S}\,\partial {\cal T}} =
\sum_{n=-\infty}^{\infty}\,\frac{\lambda/4\pi^2}{({\cal S-T}+n\beta/L)^2+L^2}
\Gamma_E({\cal S,T})\,\,\,,
\label{1}
\ee
and the magnetic kernel through

\be
\frac{\partial^2\Gamma_M}{\partial {\cal S}\,\partial {\cal T}} = 
\sum_{n=-\infty}^{\infty}\,\frac{\lambda/4\pi^2}{(n\beta)^2+ ({\cal S-T})^2+L^2}
\Gamma_M({\cal S,T})\,\,\,.
\label{2}
\ee
Both kernels are subject to the boundary conditions

\be
\Gamma_{E,M}({\cal S},0) =\Gamma_{E,M}(0,{\cal T}) =1\,\,.
\label{3}
\ee

As in the zero temperature case~\cite{zarembo} the equations are
separable in $x=({\cal S-T})/L$ and $y=({\cal S+T})/L$ with the appropriate 
boundary conditions. The temporal and spatial kernels separate

\be
\Gamma_{E,M} (x,y) =\sum_{m}\,{\bf C}_m \gamma_{E,M}^m (x)\,e^{\omega_{E,S}^m y/2}
\label{4}
\ee
with the ${\bf C}$'s fixed by the boundary conditions, and the
$\gamma$'s obeying the one-dimensional Schrodinger equation in
a periodic potential, which is

\be
\left(-\frac{d^2}{dx^2} -
\sum_{n=-\infty}^{\infty}\,\frac{\lambda/4\pi^2}{(x+n\beta/L)^2+1} \right)
\,\gamma_E^m (x) \nonumber\\
= -\frac {\omega^{m\,2}_{E}}{4}\,\gamma_E^m (x)
\label{5}
\ee
for the temporal kernel and

\be
\left(-\frac{d^2}{dx^2} -
\sum_{n=-\infty}^{\infty}\,\frac{\lambda/4\pi^2}{(n\beta/L)^2+x^2+1} \right)
\,\gamma_M^m (x) \nonumber\\
= -\frac {\omega_M^{m\,2}}{4}\,\gamma_M^m (x)
\label{5b}
\ee
for the spatial kernel. The sum of the ladder diagrams grows
exponentially: $\Gamma (0,y)\approx e^{\omega_0y/2}=e^{\omega_0T/L}$.
From (\ref{0}) it follows that the screening at finite temperature 
is given by the ground state eigenvalue 
$\alpha_{E,M}=\omega^0_{E,M} (\lambda, \beta/L)$.

In the strong coupling limit, the potential is peaked at $x=0$ modulo
$\beta$. Although the solutions to the periodic problem yields in
general a band-structure, for $\lambda\gg 1$ the ground state is just
a particle trapped at $x=0$. The ground state energy follows from the
harmonic approximation much like at zero temperature. The result for
the electric screening is 

\be
\alpha_E =
\frac{\sqrt{\lambda}}{\pi}\sqrt{\Sigma_1}-\sqrt{\frac{\Sigma_2}{2\Sigma_1}}
+{\cal O} \left( \frac 1{\sqrt{\lambda}}\right) 
\label{6}
\ee
with 

\be
\Sigma_1=&&\sum_n\frac 1{(n\beta/L)^2+1}=(\pi\,L/\beta) \coth(\pi L/\beta) \nonumber
\\
\Sigma_2=&&\sum_n\frac {1-(n\beta/L)^2}{((n\beta/L)^2+1)^3} \nonumber \\
&&=-{1\over
  2} (\pi L/\beta)^3 \coth(\pi  L/\beta)(1-\coth^2(\pi  L/\beta))\nonumber \\
&&+{1\over 4}\coth(\pi
L/\beta)-{1\over 4}(\pi L/\beta)^2(1-\coth^2(\pi L/\beta)) \,\,.
\label{7}
\ee
For the magnetic screening, the result is

\be
\alpha_M =
\frac{\sqrt{\lambda}}{\pi}\sqrt{\Sigma_1}-\sqrt{\frac{\Sigma_3}{2\Sigma_1}}
+{\cal O} \left( \frac 1{\sqrt{\lambda}}\right) 
\label{6b}
\ee
with 

\be
\Sigma_3=&&\sum_n\frac {1}{((n\beta/L)^2+1)^2} \nonumber \\
=&&-{1\over
  2} (\pi L/\beta)^2 (1-\coth^2(\pi L/\beta)) \nonumber \\
&&+{1\over
  2} (\pi L/\beta)\coth(\pi L/\beta) \,\,.
\label{7b}
\ee

In the strong coupling and to leading order, the planar resummations
yield the same screening for the electric and magnetic Wilson loops,

\be
\alpha_E\approx \alpha_M\approx
\frac{\sqrt{\lambda}}{\pi}\,\left( 1+2\sum_{n=1}^{\infty} 
\frac 1{(n\beta/L)^2+1}\right)\,\,,
\label{8}
\ee
where we have explicitly separated the thermal effects.
 
The result (\ref{8}) appears to show that in strong coupling the
potential is $stronger$ at nonzero $T$ than in vacuum. This agrees
with strong coupling Debye screening calculations~\cite{Rey_etal}.
It may appear that it contradicts the fact that the screened potential 
should vanish at large $r$. The paradox is explained by the presence
of a negative mass renormalization constant (static-charge interacting
with its Debye cloud) as mentioned at the beginning of this section.

\section{Screened ladders at finite $T$}

The static charges wich interactions we study are not
part of the  ${\cal N}$=4 YM theory, so we are free to 
give them any interactions we want. So far we have followed 
the standard notations, in which the scalar and gluon exchanges 
are equal for $\bar q q$ and cancel for
$q q$ pairs. In this case the linear divergences cancel as well.
However in-matter screening at finite $T$ is different for scalars
and gluons, so it is pertinent to discuss the screening sequentially.

\subsection{Screened Scalar Ladders}

The scalar polarization is simpler. In the lowest order it is given by
a well known bubble diagram, leading to a momentum-independent
polarization operator $m_D^2\approx\lambda T^2$. Its insertions
induce a potential between two  charges, in Euclidean space after 
analytical continuation with Matsubara periodicity:

\be
\Delta (t, {\bf x}) = &&
iT\sum_n\int\,\frac{d{\bf k}}{(2\pi)^3}\,
\frac{e^{-i\omega_n\,t +i{\bf k}\cdot{\bf x}}}{K^2+\Pi_{00}}
\label{EL9}
\ee
where $K^2=\omega_n^2+k^2$.
The zero mode $n=0$ contribution is
a standard screened potential

\be
\frac{iT}{4\pi x}e^{-m_D x}
\nonumber
\ee
and is therefore exponentially suppressed in strong coupling. 

In strong coupling, even static
charges communicate via high-frequency
quanta (short times) so we now consider 
the non-zero mode contributions in (\ref{EL6}). 
Performing the momentum integral yields 

\be
\frac{iT}{2\pi x}
\sum_{n=1} {\rm cos}(\omega_n\,t)\frac{\omega_n^2}{\omega_n^2+M_D^2}\,
e^{-\omega_n^2x/\sqrt{\omega_n^2+m_D^2}}\nonumber\,\,.
\label{EL10}
\ee
For large $\lambda$ the Debye mass $m_D\gg \omega_n$, and
the screened scalar propagator can be further simplified

\be
\Delta (t, {\bf x}) \approx\frac {iT}{4\pi x}
\left(e^{-m_Dx}+2\sum_{n=1}^\infty{\rm cos}(\omega_n t)\,
\frac{\omega_n^2}{m_D^2}\,e^{-\omega_n^2\,x/m_D}\right)\,\,.
\nonumber\\
\label{EL11}
\ee
Substituting the sum over $n$ by the integral, combining the cosine
with the exponent and completing the square in the exponent, one finally finds
that the non-static part of the induced scalar potential vanishes in the
strong coupling as  $1/\sqrt{m_D}\approx 1/\lambda^{1/4}$.

\subsection{Screened electric ladders}

We now consider exchanges of electric gluons.
In covariant gauges, the  electric propagator in Euclidean 
momentum space reads in general 

\be
\Delta_{00} (\omega, {\bf k}) = \frac 1{K^4}\left(\frac{k^4}{k^2-\Pi_{00}} +
(1-\alpha)\,{\omega^2}\right)\,\,,
\label{EL1}
\ee
where $\alpha$ is the gauge parameter and, as before, $K^2=\omega^2+k^2$.
 In configuration space the propagator is then
\be
\Delta_{00}(t,{\bf x}) =iT\sum_{n=-\infty}^{\infty}
\int\frac{d{\bf k}}{(2\pi)^3}\,e^{-i\omega_n\,t+i{\bf k}\cdot {\bf x}}\,
\Delta_{00} (\omega_n, {\bf k})\,\,.
\label{EL2}
\ee
In weak coupling we need only the static limit with $\omega_n=0$.
Therefore the gauge-sensitive term disappears and  the first
term yields the familiar Debye form with 
$\Pi_{00}(0,{\bf k})\approx m_D^2$.

In strong coupling we need the opposite limit, in which
the frequency is high with $\omega\approx \lambda^{1/4} /L \gg k\approx 1/L$.
Now the (gauge dependent) longitudinal part contributes $\Delta_{00}\approx
(1-\alpha)/\omega^2$. The $00$ part depends on the polarization
operator, which at high frequency has the generic form

\be 
\Pi_{00} (\omega, {\bf k}) \approx \lambda\,T^2\frac{k^2}{\omega^2}\,\,.
\label{pi00}
\ee
This result is the same as the one derived from hard thermal loops 
although in the opposite limit $T\gg \omega$. Below we explain why this
similarity is not fortuisious.  Inserting (\ref{pi00}) into (\ref{EL1}) 
one finds that one power of $k^2$ can be canceled, and the first term 
becomes of order $(k^2/\omega^2)/(\omega^2 + \omega_p^2)$,
with the denominator capable of producing the plasmon pole. 
However, this term is clearly subleading 
as compared to the longitudinal one. This happens because it
lacks the enhancement through $\omega^2$ in the numerator.

Inserting the electric part (\ref{loop}) in (\ref{EL2}) and
performing the momentum integration yields 

\be
\Delta_{00} (t, {\bf x}) =&&
\frac {iT}4(\frac{e^{-m_Dx}}{\pi x} 
+2\sum_{n\neq 0}^\infty{\rm
cos}(\omega_n\,t)\,e^{-\omega_nt}\,\frac{\omega_n^2}{\omega_n^2+\omega_p^2}\nonumber\\
&&\times\left(\frac 1{\pi x}+\frac
1{2\pi\omega_n}((1-\alpha)(\omega_n^2+\omega_p^2)-\omega_n^2)\right))
\nonumber\\
\label{EL3}
\ee
where the first and screened contribution arises from the zero mode
and the second contribution arises from the non-zero modes in the high
frequency (short time) regime $\omega\gg k$. In the strong coupling
limit $\omega_D/T\approx \sqrt{\lambda}\gg 1$ (\ref{EL3}) reduces to

\be
\Delta_{00} (t, {\bf x})
=(1-\alpha)\frac{iT}{4\pi}\sum_{n=1}^\infty
{\rm cos}(\omega_n\,t)\,\omega_n\,e^{-\omega_n\,x}
\label{EL4}
\ee
which is controlled by the first Matsubara frequency at large
distance. It is readily checked that for $t=0$ 

\be
\Delta_{00}(0, {\bf x})=(1-\alpha)\frac {ix\,T^2}{8\,{\rm sh}^2\pi\,T\,x}
\label{EL5}
\ee
which reduces to the free electric part of the covariant propagator
at $T=0$. This part is the chief contribution to the screened electric
ladders in the strong coupling. Indeed, a rerun of the previous 
Bethe-Salpeter resummation using (\ref{EL4}) as the screened ladder 
shows that only the short time limit of (\ref{EL5}) contributes as
$\lambda\gg 1$. In particular, the electric potential
between two Wilson lines with only adjoint $A$ fields is

\be
V_E (Tb, \alpha) = -\frac{\sqrt{\lambda}/\pi}{b}\,{\bf F}_E\,(bT, \alpha)
\label{EL5b}
\ee
with

\be
{\bf F}_E(bT, \alpha) =\sqrt{1-\alpha}\,\frac{\pi\,T\,x}{|{\rm
sh}(\pi\,T\,x)|}\,\,,
\label{EL6}
\ee
which is seen to reduce to the $T=0$ result in Feynman gauge. 
As noted above, the result (\ref{EL6}) yields a screening length
of order $1/(\pi\,T)$ in strong coupling, which is {\it larger}
than the $1/(\sqrt{\lambda} T)$ expected from weak coupling. Again,
the strong coupling result is even consistent with current QCD 
lattice simulations which suggest that the Debye screened 
potential screens at about a distance of order  $1/\pi T$.

Finally, the result (\ref{EL6}) is gauge sensitive, and one may
question the interpretation of our results. Of course the same
criticism for summing the ordered ladders hold at $T=0$ as well.
However, we expect our main observations in covariant gauge to hold 
for the reason that the results agree overall with results from the
ADS/CFT correspondence both at $T=0$ and finite $T$ for the electric
sector. Retardation is best captured in covariant gauges, and that was
seen as key in describing the screening at work in the ground state.

\subsection{Hard thermal loops at short times}

The effects of matter is more than the Bose enhancement.  It is
usually due to a genuine screening of the Coulomb 
interaction. How would this work at strong coupling? The answer
would be in general hopeless, except for yet another important
observation: The induced interaction occurs in our case over
very {\it short} times. This means that the we probe the thermalized but
strongly coupled Coulomb phase over short period of times, in which
case the thermal distributions are left unchanged. As a result, the
high frequency modes are screened at strong coupling in exactly the
same way as they are in {\it weak} coupling at high temperature. 

This is best seen using a first quantized analysis of screening
in terms of transport arguments, whereby the 
only assumption needed to derive the hard thermal loops is that
the thermal distribution be expandable into $f=f^0+f^1\approx f^0$
for $f^1/f^0\ll 1$ at short times. 
In Euclidean space (Matsubara frequencies) the hard thermal loops
reduce to

\be
\Pi_{00} (\omega, k) = && -\omega_D^2\left(1-\frac{\omega}{2k}\,
{\rm ln}\left|\frac{\omega+k}{\omega-k}\right|\right)\nonumber\\
\Pi_{ii} (\omega, k) = && \frac{\omega_D^2}2\left(\frac{\omega}{k}\,
{\rm ln}\left|\frac{\omega+k}{\omega-k}\right|\right)\,\,.
\label{loop}
\ee
The zero Matsubara frequencies are electrically screened 
with $\Pi_{00} = -\omega_D^2$ and magnetically free with
$\Pi_{ii}=0$. The situation is somehow reversed at high 
Matsubara frequencies with

\be
&&\Pi_{00}\approx-\omega_p^2\,\frac{k^2}{\omega^2}\nonumber\\
&&\Pi_{T} (\omega, k) \approx  \omega_D^2 
\label{loopb}
\ee
The plasmon frequency is defined as $\omega_p^2=\omega_D^2/3$ with the
Debye frequency  $\omega_D=m_{G,\phi,F}$ for the gluons, scalars and
quarks read respectively

\be
m_G^2\approx m_F^2\approx m_\Phi^2\approx \lambda T^2\,\,.
\ee
The electric modes are free at high
frequency, while the magnetic modes are Debye screened. This is 
just the opposite of what happens at low frequency. Still, {\it both}
the dielectric constant and inverse magnetic permittivity 
{vanish} at large frequency. The thermal medium is transparent
to high frequency electric and magnetic fields as it should.

\subsection{Screened Magnetic Ladders}

The real insertions (\ref{loop}) induce a new potential between
two moving charges in Euclidean space after analytical continuation.
In covariant gauge, the resummed magnetic propagator in Euclidean 
momentum space reads ($K^2=\omega^2+k^2$)

\be
\Delta_{ij} (t, {\bf x}) = &&(-\delta_{ij} + \nabla_i\nabla_j)\,
\nonumber\\
&&\times iT\sum_n\int\,\frac{d{\bf k}}{(2\pi)^3}\,
\frac{e^{-i\omega\,t +i{\bf k}\cdot{\bf x}}}{K^2+\Pi_{T}}
\nonumber\\
&&+\nabla_i\nabla_j\,
iT\sum_n\int\,\frac{d{\bf k}}{(2\pi)^3}\,
\frac{e^{-i\omega\,t +i{\bf k}\cdot{\bf x}}}{K^4}
\nonumber\\
&&\times \left(-\frac{\omega^2}{k^2+\Pi_{00}}+(1-\alpha)\right)\,\,.
\label{EL7}
\ee
Performing the momentum integrations, taking the strong coupling
limit $\lambda\gg 1$ and contracting the answer with the particle
velocities ${\bf v}_{1,2}$ attached to the external Wilson lines
yield

\be
&&{\bf v}_{1i}{\bf v}_{2j}\,\Delta_{ij}=
2{\bf v}_1\cdot{\bf v}_2\,\frac{iT}{4\pi\,x^3}\nonumber\\
&&\times\left(-\frac{x^2}2\,\delta_{=0}+\frac 1{m_M^2} \delta_{\neq 0} 
+\frac 1{\omega_p^2}-\frac 1{\omega_D^2}\right)
\label{EL8}
\ee
where we have inserted a magnetic mass $m_M/T\approx \lambda$ on the
transverse magnetic zero mode. $\delta_{=0}$ is the contribution when 
the magnetic mass is zero (long range field), while $\delta_{\neq 0}$ 
is the contribution when the magnetic mass is inserted (short range
field). Clearly, the magnetic contribution survives the strong coupling
limit only in the absence of a magnetic mass. The contribution to the 
ladder resummation follows from the basic insertion

\be
\frac {i\lambda}{2}\,{\bf v}_{1i}{\bf v}_{2j}\,\Delta_{ij}\,\,.
\nonumber
\ee

The screened magnetic ladders yield a magnetic
potential that is different from the electric one derived above in
coupling and range if a magnetic mass is present. In the absence of
an induced magnetic mass, the electric and magnetic potentials 
exhibit a similar behavior in coupling, but are distinct in range.
The magnetic potential still has infinite range, while the electric
potential has a finite range of order $1/\pi T$ as we indicated above.
The simple AdS/CFT argument provided earlier leads to the same electric
and magnetic screening length. The understanding of this point from the
gauge theory standpoint is worth pursuing.

\section{Summary and outlook}

We have exploited the fact that ${\cal N}=4$ SYM is in a Coulomb
phase at all couplings, to argue that in strong coupling (Maldacena
regime) color charges only communicate over very short period of times
$t\approx L/\lambda^{1/4}$ for a fixed separation $L$. This physical
observation is enough to show why ladder-like diagrams in the gauge
theory reproduce the modified Coulomb's law obtained by the AdS/CFT 
correspondence using classical gravity. This class of diagrams is
however not complete since the answer is gauge-sensitive. We have
shown that the insertion of additional quanta along the ladder
brings about extra factors of $\lambda\int\,d^4{\bf x}\approx \lambda^0$,
leading to a quasi-string geometry. In ${\cal N}=4$ SYM theory the
quasi-string has a transverse to a longitudinal ratio of order 
$(L/\lambda^{1/4})/L\ll 1$, but is not confining.

These observations are generic and suggest that in the gauge theory the
modified Coulomb potential applies equally well to relativistic and
non-relativistic charges. Indeed, since the relativistic particles move
with velocity $v\approx 1$, the color reordering encoded in the 
modified Coulomb potential goes even faster through a virtual quantum 
exchange with velocity $v\approx \lambda^{1/4}\gg 1$. {\it At strong
coupling and/or large number of colors}, the  charge is so large that 
color rerrangement is
so prohibitive unless it is carried instantaneously. This is the only 
way Coulomb's law could budget its energy. The same observations extend 
to finite temperature
where we have shown that the modified Coulomb potential acquires a
screening length of the order of $1/T$ irrespective of how strong is
the coupling. This observation seems to be consistent with current
lattice simulation of non-conformal gauge theories in their moderatly
strong Coulomb phase, e.g. QCD.

We have analyzed the effects of a (supercritical) Coulomb field on the 
motion of colored relativistic particles. Bound states form whenever 
the squared Coulomb potential balances the effects of centrifugation.
In strong coupling the resulting bound state spectrum is oscillator-like,
in agreement with the tower of resonances observed using the AdS/CFT 
correspondence~\cite{Starinets,Teaney}. Rather unexpectedly, we have
found that even though the trajectory of any particular Coulomb bound
state depends critically on the coupling $\lambda$, their average
density remains constant in the  strong coupling
domain. This finally leads to a {\em universal
gas of composites} in strongly coupled CFT, a nice parallell with recent
developments in QCD~\cite{SZ_newqgp}.

Clearly our work is only the first attempt in trying to understand
from the gauge theory standpoint the intricate and surprising results
obtained by the AdS/CFT correspondence both in vacuum and matter. 
In particular, our discussion of the spectroscopy
of composites, Debye screening, thermodynamics and kinetics
of matter were rather sketchy, with only the qualitative trends emphasized.
All of this can obviously be worked out with more details that go
beyond the scope of the current presentation.

However, it is clear that the dynamical picture we have put forward
from the gauge theory standpoint goes beyond the confines of
supersymmetry or the conformal nature of the strongly coupled gauge
theory considered here. Indeed, we believe that our results extend
to all gauge theories in their Coulomb phase at strong coupling, 
thereby opening a window of understanding to a variety of physical
problems in different settings.

\vskip 1.25cm
{\bf Acknowledgments}
\\\\
We thank S.J. Rey for a stimulating discussion at the beginning of
this work, explaining to us the details of his published results. 
This work was supported in parts by the US-DOE grant
DE-FG-88ER40388.

\end{document}